\documentclass[conference]{IEEEtran}

\usepackage{cite}
\usepackage{amsmath,amssymb,amsfonts}
\usepackage{algorithmic}
\usepackage{graphicx}
\usepackage{textcomp}
\usepackage{xcolor}
\usepackage{hyperref}
\def\BibTeX{{\rm B\kern-.05em{\sc i\kern-.025em b}\kern-.08em
    T\kern-.1667em\lower.7ex\hbox{E}\kern-.125emX}}

\begin{document}

\title{Structural Cellular Hash Chemistry}

\author{
\IEEEauthorblockN{Hiroki Sayama}
\IEEEauthorblockA{\textit{Binghamton Center of Complex Systems, Binghamton University, State University of New York}, 
Binghamton, NY, USA\\
\textit{Waseda Innovation Lab, Waseda University}, Tokyo, Japan\\
sayama@binghamton.edu \quad ORCID: 0000-0002-2670-5864}
}
\maketitle

\begin{abstract}
Hash Chemistry, a minimalistic artificial chemistry model of open-ended evolution, has recently been extended to non-spatial and cellular versions. The non-spatial version successfully demonstrated continuous adaptation and unbounded growth of complexity (size) of self-replicating entities, but it did not simulate multiscale ecological interactions among the entities. On the contrary, the cellular version explicitly represented multiscale spatial ecological interactions among evolving patterns, yet it failed to show meaningful adaptive evolution or complexity growth. It remains an open question whether it is possible to create a similar minimalistic evolutionary system that can exhibit all of those desired properties at once, within a computationally efficient framework. Here we propose an improved version of Cellular Hash Chemistry, called ``Structural Cellular Hash Chemistry'' (SCHC). In SCHC, individual identities of evolving patterns are explicitly represented and processed as the connected components of the nearest neighbor graph of active (non-empty) cells. The neighborhood connections are established by connecting active cells with other active cells in their Moore neighborhoods in a 2D cellular grid. Evolutionary dynamics in SCHC are simulated via pairwise competitions of two randomly selected patterns, following the approach used in the non-spatial Hash Chemistry. SCHC's computational cost was significantly less than the original and non-spatial versions. Numerical simulations showed that these model modifications achieved spontaneous movement, self-replication and unbounded growth of complexity (size) of spatial evolving patterns, which were clearly visible in space in a highly intuitive manner. Detailed analysis of simulation results showed that there were spatial ecological interactions among self-replicating patterns and their diversity was also substantially promoted in SCHC, neither of which was present in the non-spatial version. 
\end{abstract}


\section{Introduction}

Open-endedness is one of the most actively discussed topics in the fields of Artificial Life and Artificial Intelligence \cite{stanley2019open,
packard2019overview,stepney2021modelling,borg2023evolved,stepney2023open} and particularly in the distributed dynamical systems-based ALife research community \cite{adams2017,chan2023,nichele2024special,sayama2024review}. To facilitate open-ended evolution in such ALife models, we proposed the concept of ``cardinality leap'', i.e., a general mathematical operation to drastically increase the cardinality of the possibility space by allowing the formation of higher-order entities. Its effectiveness was demonstrated in ``Hash Chemistry'' \cite{sayama2019}, a spatial artificial chemistry model in which a hash function was used just as an example of a convenient ``oracle'' mechanism to evaluate the fitness of higher-order entities of any size/scale. The original Hash Chemistry was computationally very expensive, and it was also somewhat bounded in terms of the complexity growth of self-replicating entities. 

Recently, we proposed a non-spatial variant of Hash Chemistry based on multisets \cite{sayama2024a} that achieved substantial speed-up of model simulations and significantly better demonstration of continuous adaptation and unbounded growth of complexity of self-replicating entities\footnote{Note that, in Hash Chemistry and its variants, the complexity of self-replicating entities is characterized merely by the size of self-replicators because of the extreme simplicity of these models. Characterization of other forms of complexity would likely require more complex evolutionary models.}. However, this non-spatial version was created at the cost of nontrivial spatial ecological interactions of self-replicating higher-order entities. To address this limitation, we also created a prototype of ``Cellular Hash Chemistry'' \cite{sayama2024b} by placing components of self-replicating entities in a 2D cellular grid. This explicit spatial representation allowed for multiscale spatial ecological interactions among evolving patterns, and its simplified updating rules also reduced computational costs drastically. However, this prototype cellular version failed to show meaningful adaptive evolution or complexity growth of self-replicating entities. It has thus remained an open question whether it would ever be possible to create a similar minimalistic evolutionary system that could exhibit all of those desired properties at once, within a computationally efficient framework. 

To address this open question, here we propose a new improved version of Cellular Hash Chemistry, called ``Structural Cellular Hash Chemistry'' (SCHC). In SCHC, individual identities of evolving patterns are explicitly represented and processed as the connected components of the nearest neighbor graph of active (non-empty) cells. The neighborhood connections are established by connecting active cells with other active cells in their Moore neighborhoods in a 2D cellular grid. Evolutionary dynamics in SCHC are simulated via pairwise competitions of two randomly selected individuals (connected components), following the approach used in the non-spatial Hash Chemistry \cite{sayama2024a}. Numerical simulations showed that these model modifications achieved spontaneous movement, self-replication and unbounded growth of complexity of spatial evolving patterns in SCHC. Major improvements from the non-spatial version include that these evolutionary behaviors were clearly visible in space in a highly intuitive manner; there were nontrivial spatial ecological interactions among self-replicating patterns; and the diversity of those patterns was substantially promoted. Moreover, SCHC's computational cost was significantly reduced compared to the original version and even to the non-spatial version, achieving all of the desired properties (continuous adaptation, unbounded growth of complexity, spatial ecological interactions among self-replicating entities, diversity of populations, and computational efficiency) within a single model framework.  

The codes for simulation and data analysis/visualization are available from the author's GitHub page\footnote{\url{https://github.com/hsayama/Structural-Cellular-Hash-Chemistry}}.

\begin{figure*}
\centering
\fbox{\includegraphics[width=0.23\textwidth]{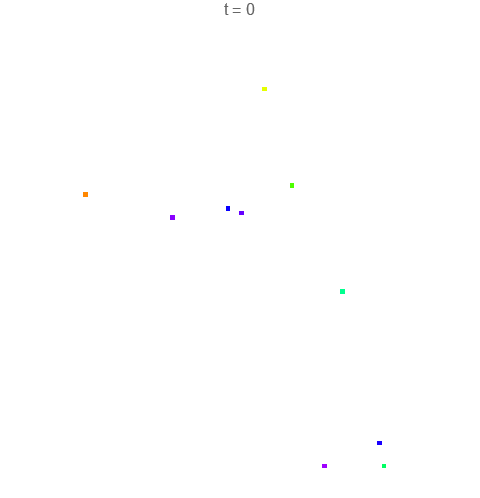}}
\fbox{\includegraphics[width=0.23\textwidth]{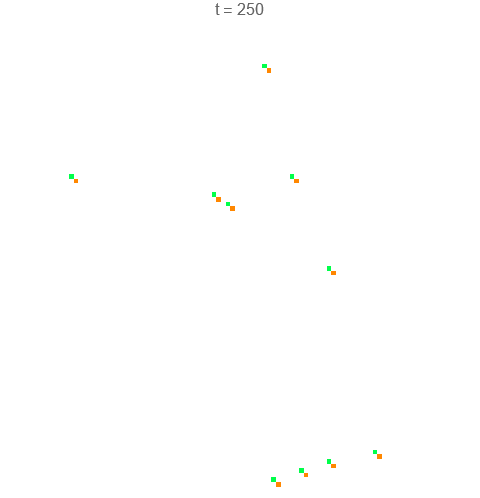}}
\fbox{\includegraphics[width=0.23\textwidth]{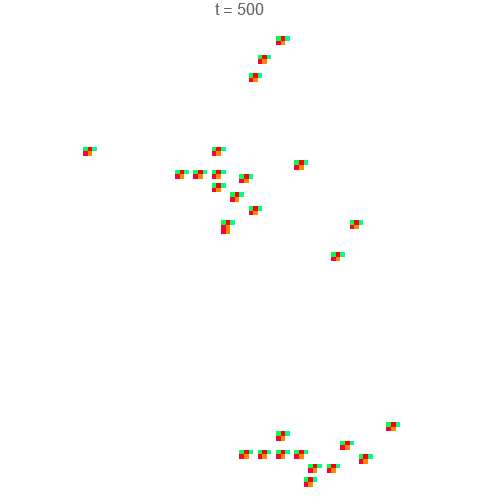}}
\fbox{\includegraphics[width=0.23\textwidth]{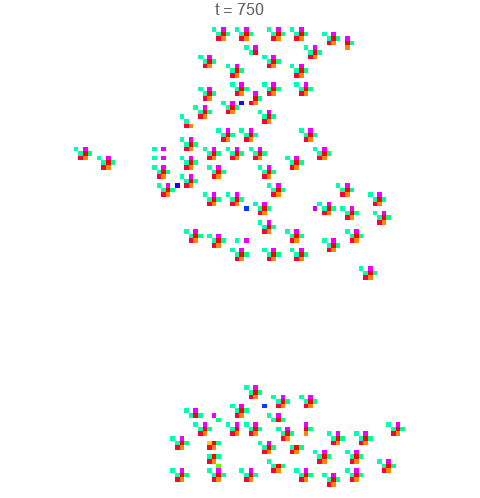}}\\~\\
\fbox{\includegraphics[width=0.23\textwidth]{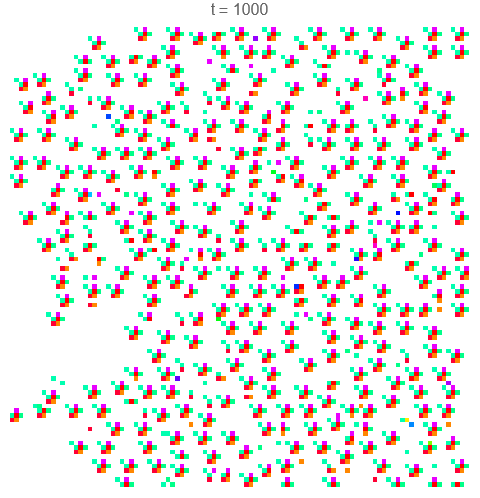}}
\fbox{\includegraphics[width=0.23\textwidth]{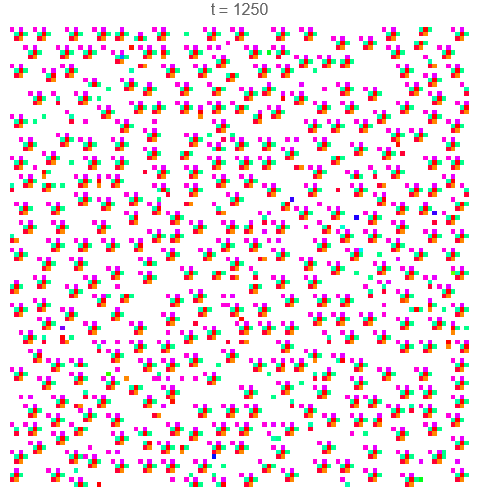}}
\fbox{\includegraphics[width=0.23\textwidth]{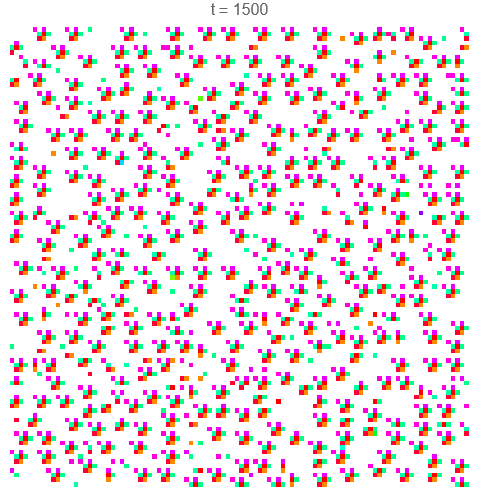}}
%\fbox{\includegraphics[width=0.25\textwidth]{1750.png}}
\fbox{\includegraphics[width=0.23\textwidth]{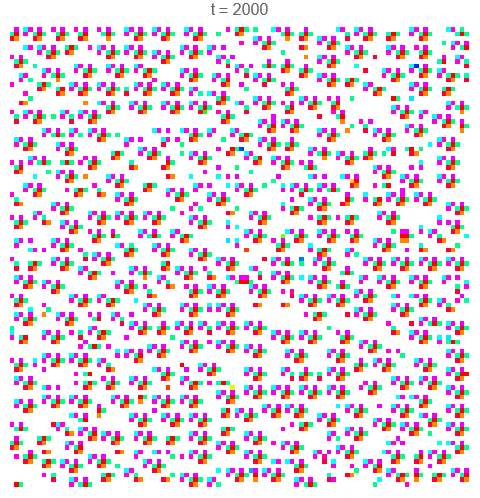}}
\caption{A sample simulation run of Structural Cellular Hash Chemistry. Snapshots of system configurations are arranged temporally from left to right and then top to bottom ($t = 0, 250, 500, 750, 1000, 1250, 1500, 2000$). Colors represent different element types, and blank (white) spaces represent empty cells. It is observed in these visualizations that the self-replicating patterns gradually proliferate and evolved to larger forms with more complex nontrivial structures (also see Fig.\ \ref{size-growth-example}).}
\label{sample-run}
\end{figure*}

\begin{figure}
\centering
\includegraphics[width=\columnwidth]{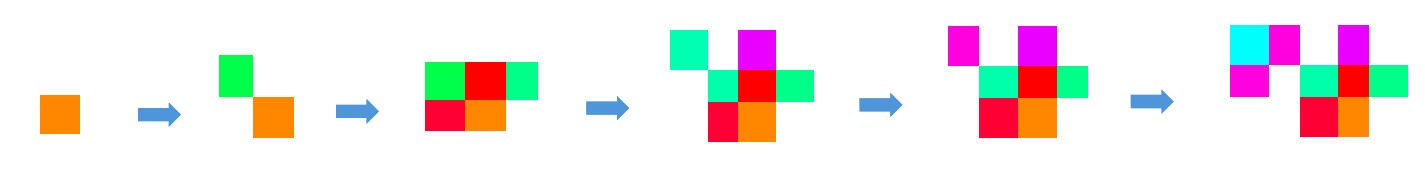}
\caption{Example of spontaneous growth of complexity of self-replicating patterns (extracted from the simulation run shown in Fig.\ \ref{sample-run}).}
\label{size-growth-example}
\end{figure}

\section{Model}
\label{model}

In Structural Cellular Hash Chemistry (SCHC), we assume the same cellular automata-like $L \times L$ grid as used in \cite{sayama2024b}, where each cell can be empty or contain an individual element of type $i$ ($i \in \{1, \ldots, k\}$). In this study, we used $L = 100$, $k = 1000$, and initial configurations that were almost empty but with 10 randomly generated initial individual elements scattered over space (Fig.~\ref{sample-run}, top left). These settings were also adopted from \cite{sayama2024b}.

The updating rule of SCHC is designed following the random selection and pairwise\footnote{In the original prototype of Cellular Hash Chemistry \cite{sayama2024b}, four-way competition was used. In this study, we reduced it to pairwise competition in order to make the model simpler.} competition procedure adopted in \cite{sayama2024a} and proceeds as follows:
\begin{enumerate}
\item Randomly choose two active (non-empty) cells from the space. 
\item For each of the selected active cells, find a connected component it belongs to in the nearest neighbor graph (NNG) of active cells\footnote{In the actual implementation of the simulation algorithm, NNGs for the entire space are never constructed. Rather, the list of active cells is dynamically maintained throughout the simulation, and a connected component is constructed locally through neighbor search from the selected cell every time an active cell is randomly chosen.}. The NNG is constructed by connecting active cells that are adjacent within the Moore neighborhood of each other.\footnote{It is possible that the detected two connected components may be the same one, but this does not cause any problem in the simulation process.} Cut-off boundary conditions were assumed at the borders of the space.
\item Convert each connected component into a sorted list of {\tt [(x, y), c ]}, where {\tt (x, y)} is the relative coordinate of an active cell within the connected component and {\tt c} is the type of that cell. The origin of the relative coordinate is set to the top-left corner of a bounding rectangle that circumscribes the connected component.
\item Calculate the fitnesses of the two sorted lists obtained above using a hash function. Specifically, the fitness value is calculated as $(h(\mathrm{ls}) \; \mathrm{mod} \; M) / M$, where $h(x)$ is the hash function\footnote{We used Mathematica 14's {\tt Hash} function in its default settings.}, $\mathrm{ls}$ is the sorted list that describes the spatial pattern of the selected connected component, and $M$ is a normalization parameter ($M = 100000000$ in this study; the same as in \cite{sayama2024b}). Then determine the winning pattern with greater fitness between the two.
\item Let the winning pattern copy itself into the region where the other pattern existed. Specifically, a rectangular region obtained by expanding the rectangle circumscribing the winning pattern by one row/column toward the top, bottom, left, and right\footnote{This one-row/column expansion of the rectangular region is implemented to make sure the copied pattern is separated from the nearby patterns by empty cells and thus can maintain its identity. This expansion also allows for the growth of the size of the replicating patterns since cells in those expanded rows/columns are subject to mutation.} is copied to the other pattern's area so that their centers match. In this copying process, introduce random mutations with small probability\footnote{In this study, each cell within the copied pattern was reset to an empty state with 0.1\% probability or mutated to a random value selected from $\{1, \ldots, k\}$ with 0.2\% probability.}.
\end{enumerate}
The above steps are repeated until the total number of active cells contained in the winning patterns reaches the total number of active cells in the entire system, so that every active cell has a chance to replicate, on average. This constitutes one update of the entire system configuration, which is consistent with the treatments used in earlier work \cite{sayama2019,sayama2024a,sayama2024b}. We implemented a simulator of SCHC in Wolfram Mathematica 14 and conducted numerical simulations. Simulations of SCHC took more computational time than the prototype version \cite{sayama2024b} in order to capture spatial interactions among cells, but it still remained computationally efficient (e.g., a simulation for 2000 updates with $L=100$ would complete in about 15 to 20 minutes in an ordinary laptop without any code optimization or parallelization). 

\section{Results}

We conducted a series of simulations of the proposed SCHC model on a Windows 11 (64-bit) desktop workstation with an Intel i9 CPU (10 cores) at 3.70 GHz with 64 GB RAM (this is the same hardware as used in \cite{sayama2024a,sayama2024b}). Numerical simulation of SCHC successfully exhibited highly intriguing, visibly recognizable evolutionary dynamics of self-replicating patterns whose size tended to increase over time\footnote{Animations of simulation results can be seen at \url{https://www.youtube.com/@ComplexSystem/shorts}.}. Figure \ref{sample-run} presents several snapshots of the system configurations taken from a sample simulation run over time, in which one can see spatial patterns (i.e., connected components) gradually self-replicate and evolve toward larger, more complex nontrivial structures (details shown in Fig.\ \ref{size-growth-example}). 

In SCHC, spatial patterns exhibit spontaneous trembling movement and binary fission-like self-replication even though such behaviors are not explicitly implemented in the system's updating algorithm (Section \ref{model}). Spontaneous movement occurs when the original pattern has a new active cell emerging nearby due to mutation, shifting the center of the pattern slightly, and then quickly becomes overwritten by another original pattern. Binary fission-like self-replication occurs when mutations and/or spatial interactions among patterns accidentally overwrite an active cell with an empty state and thus break a single connected component into two (or more), which then quickly become overwritten by the original patterns each. These stochastic fluctuations drive the proliferation of patterns in space, as observed in the top row of Fig.\ \ref{sample-run}. Meanwhile, spontaneous growth of complexity of replicating higher-order entities is driven by the exponential relationship between the number of elements within a self-replicating higher-order entity and the cardinality of the design space (and thereby the greater chance for higher fitness values). SCHC successfully inherits this ``cardinality leap'' \cite{sayama2019} property of the Hash Chemistry model family.

We conducted a set of 100 independent simulation runs of the SCHC model for the statistical analysis of its evolutionary dynamics. During each simulation run, every single copying event was saved in a log file with the spatial structure of the copied pattern and its fitness value. There were 13 out of 100 cases in which replicating entities did not exhibit any meaningful long-term evolutionary changes (as determined by the small size of the log file; typically less than 1MB), which were excluded from data visualization and analysis. The remaining 87 log files were post-processed and analyzed for visualizations and statistical analyses, which are reported below.

\begin{figure}
\centering
\includegraphics[width=\columnwidth]{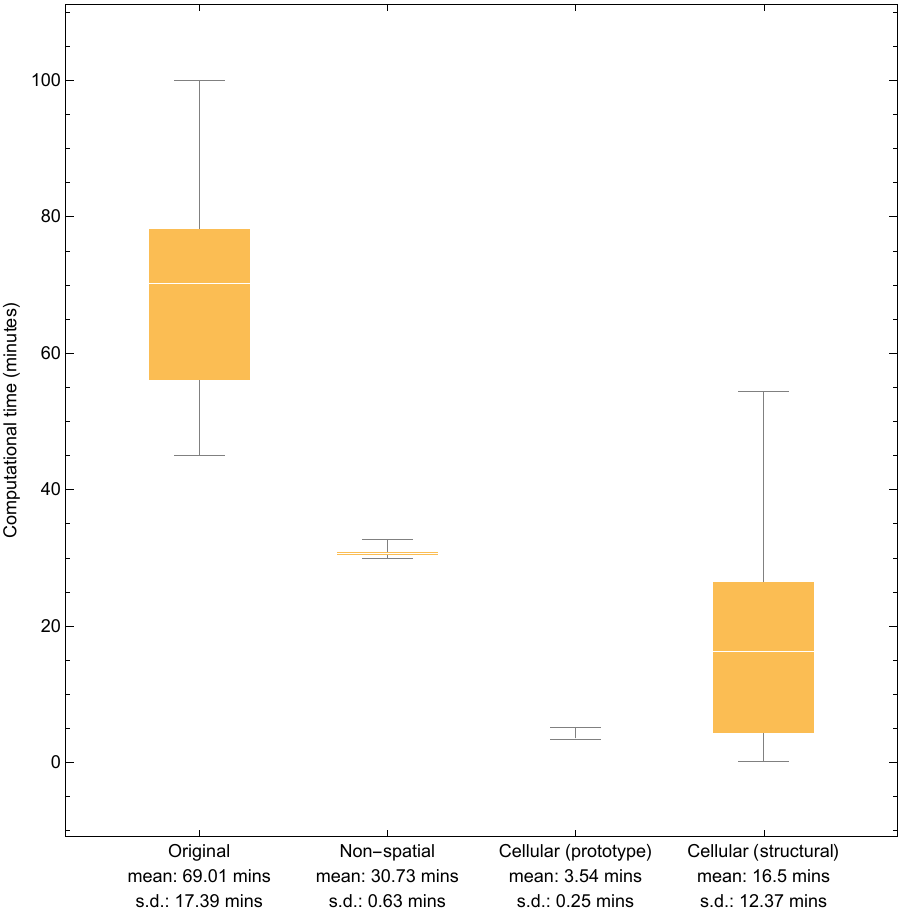}
\caption{Box-whisker plots comparing the distributions of computational time needed to complete one simulation run for 2000 iterations among the original Hash Chemistry \cite{sayama2019} (1st), the non-spatial version \cite{sayama2024a} (2nd), the prototype version of Cellular Hash Chemistry \cite{sayama2024b} (3rd), and the proposed Structural Cellular Hash Chemistry (4th). The vertical axis shows the length of simulation time in minutes on a Windows 11 (64-bit) desktop workstation with an Intel i9 CPU (10 cores) at 3.70 GHz with 64 GB RAM. The four conditions are all statistically very significantly different from each other (ANOVA; $p<10^{-72}$).}
\label{fig:performance-comparison}
\end{figure}

We first compared the computational efficiency of the proposed SCHC model with the three other versions of previously published Hash Chemistry models \cite{sayama2019,sayama2024a,sayama2024b}. Figure \ref{fig:performance-comparison} shows the distributions of the computational time needed to complete one simulation run for 2000 iterations in the four models. While this is not a rigorous comparison (as the models are quite different from each other), the practical computational efficiency of SCHC is clearly seen in Fig.\ \ref{fig:performance-comparison} in comparison with the original version (1st plot) and even with the non-spatial one (2nd plot). The mean computational time of SCHC was nearly half of the non-spatial version's, although SCHC could sometimes take much longer time especially when the evolved patterns become large and complex (as indicated by the longer whisker in the 4th plot). SCHC was significantly slower than the prototype version of the Cellular Hash Chemistry model (3rd plot), which was because of the nearest neighbor detection involved in SCHC. We consider this a necessary computational cost worth paying in order to obtain meaningful spatial evolutionary dynamics. Statistical analysis (ANOVA) showed a very strong statistically significant difference between every pair of the four conditions ($p<10^{-72}$; details of results not shown).

\begin{figure}[t!]
\centering
\includegraphics[width=\columnwidth]{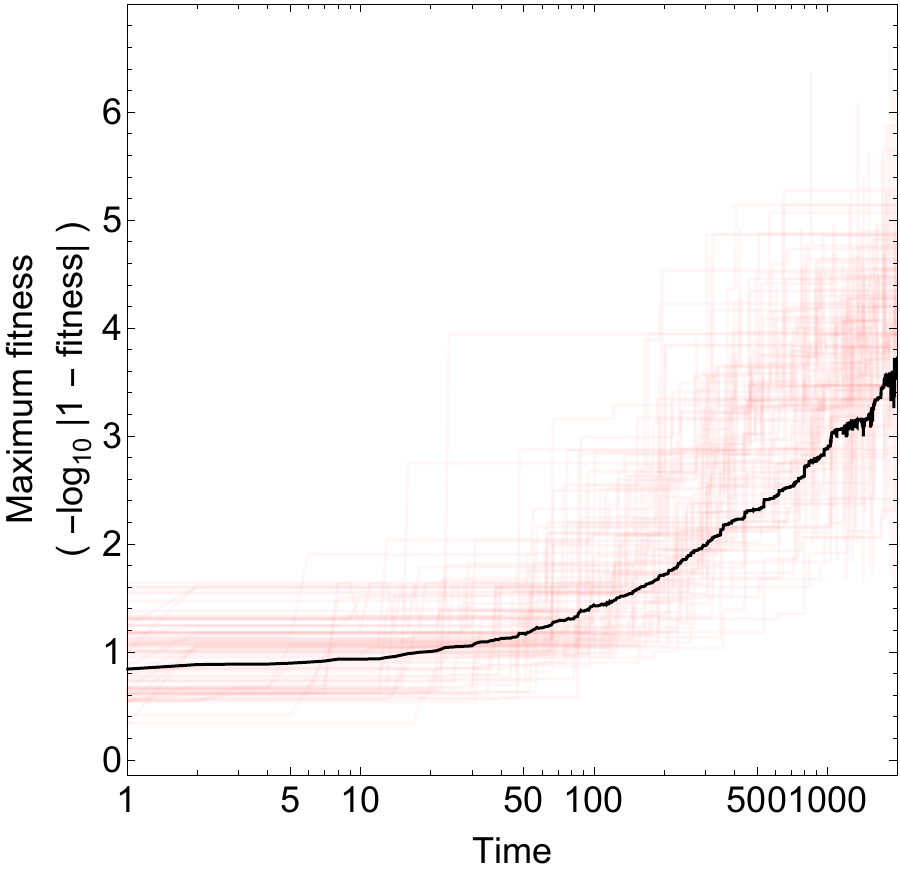}\\~\\
\includegraphics[width=\columnwidth]{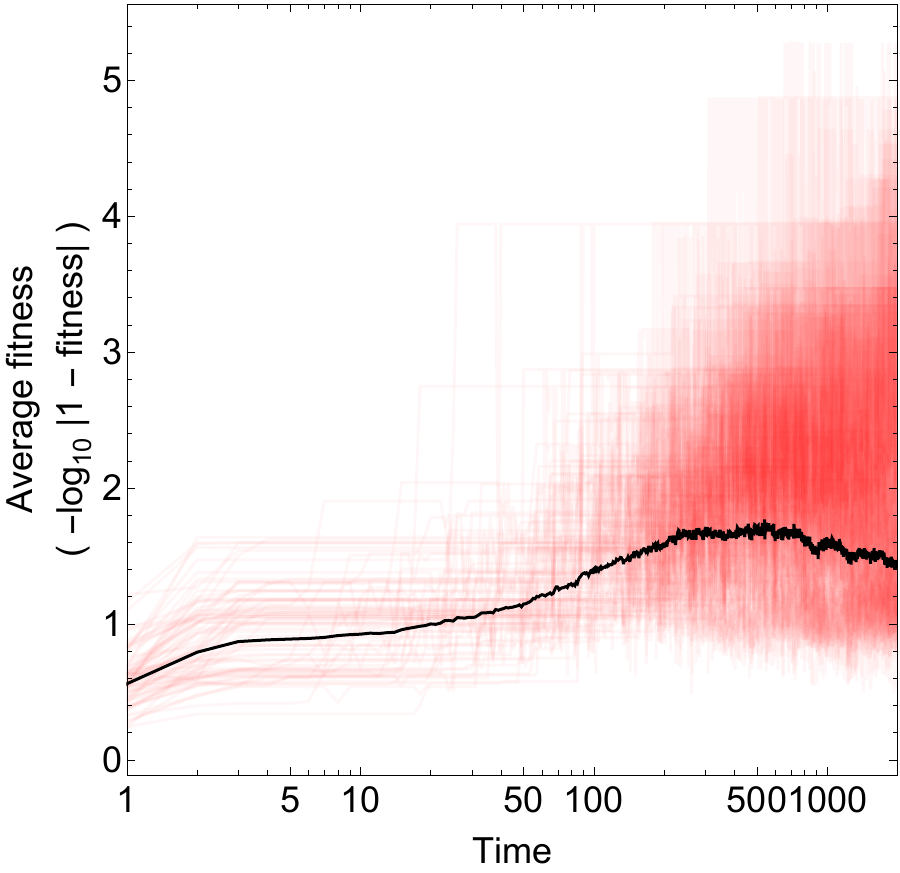}
\caption{Fitness values (i.e., return values of the hash function) of successfully replicated patterns in SCHC simulations. Top: Maximum fitness value observed in each time step. Bottom: Average fitness value in each time step. The red curves show results of 87 independent simulation runs, while the black solid curve shows their average. The time is in log scale to show long-term trends clearly. The fitness values are visualized using $-\log_{10} | 1 - \mathrm{fitness} |$ to visualize increasingly finer improvement of the fitness that progresses over the course of simulation. The maximum fitness (top) showed a continuous increase over time, whereas the average fitness (bottom) showed an initial increase followed by a slight decrease in the long run, which was not observed in the non-spatial model \cite{sayama2024a} and indicates the presence of nontrivial spatial ecological interactions among patterns. See text for more details.}
\label{fig:adaptation}
\end{figure}

Figure \ref{fig:adaptation} presents time series of the maximum fitness values (top) and the average fitness values (bottom) of successfully replicated patterns observed in each time step. Note that these ``fitness'' values are just outputs of the hash function used to choose winners in pairwise competition; they are {\em not} actual fitness (i.e., number of successfully replicated offspring per unit of time) of self-replicating patterns in ecological/evolutionary sense. In these plots, the fitness values are visualized using $-\log_{10} | 1 - \mathrm{fitness} |$ (i.e., how close, or for how many digits, the fitness approaches its theoretical maximum 1), in order to visualize increasingly finer improvement of the fitness that progresses over the whole course of simulation. The maximum fitness (Fig.\ \ref{fig:adaptation}, top) showed a continuous increase over time. In the meantime, the average fitness (Fig.\ \ref{fig:adaptation}, bottom) showed an initial increase followed by a slight decrease in the long run, which was not observed in the non-spatial model \cite{sayama2024a} and is a new finding observed in SCHC.

We note that the observed decreasing trend of the average ``fitness'' (Fig.\ \ref{fig:adaptation}, bottom) started at around $t=500-700$, which roughly corresponds to a typical time point when self-replicating patterns began to form clustered populations. In those crowded clusters, patterns do interact and interfere with each other spatially, and therefore the ``fitness'', i.e., the return value of the hash function, is no longer the only factor that determines the success of self-replication and self-maintenance of patterns. Rather, each pattern's implicit robustness against spatial interactions begins to play crucial roles in the evolutionary survival of the pattern. The decreasing ``fitness'' captured in Fig.\ \ref{fig:adaptation} thus strongly indicates the presence of nontrivial spatial ecological interaction that made the actual evolutionary fitness of patterns deviate from the hash function's return values (``fitness'' plotted in Fig.\ \ref{fig:adaptation}), which never occurred in the non-spatial model \cite{sayama2024a} where replication success was simply determined by the hash values only.

\begin{figure}[t!]
\centering
\includegraphics[width=\columnwidth]{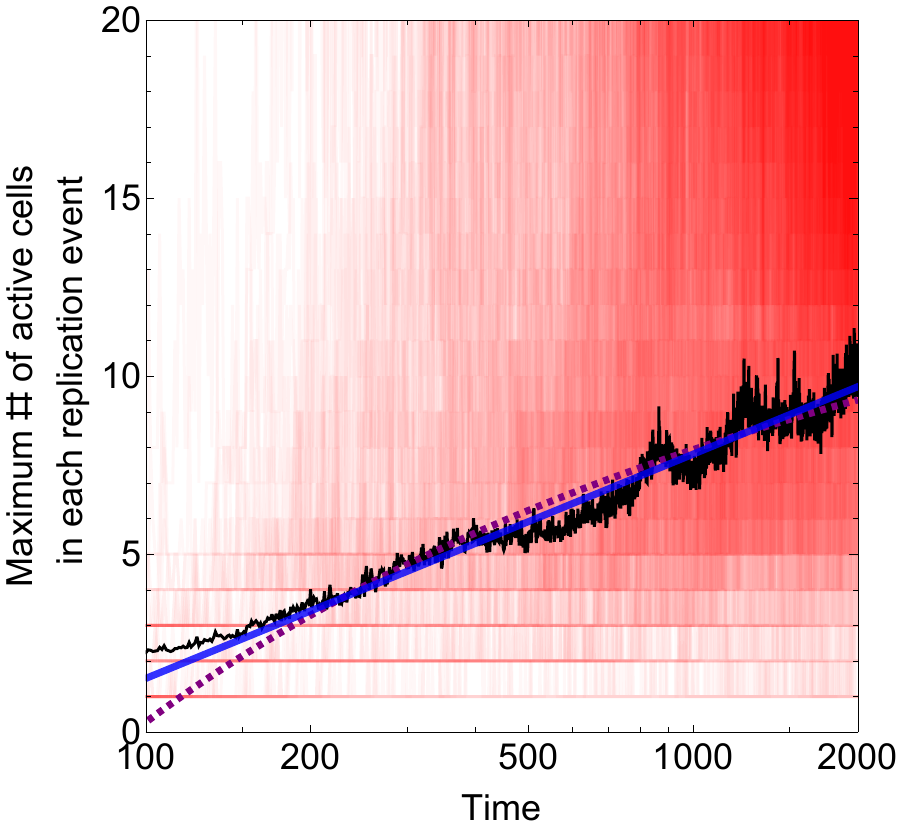}\\~\\
\includegraphics[width=\columnwidth]{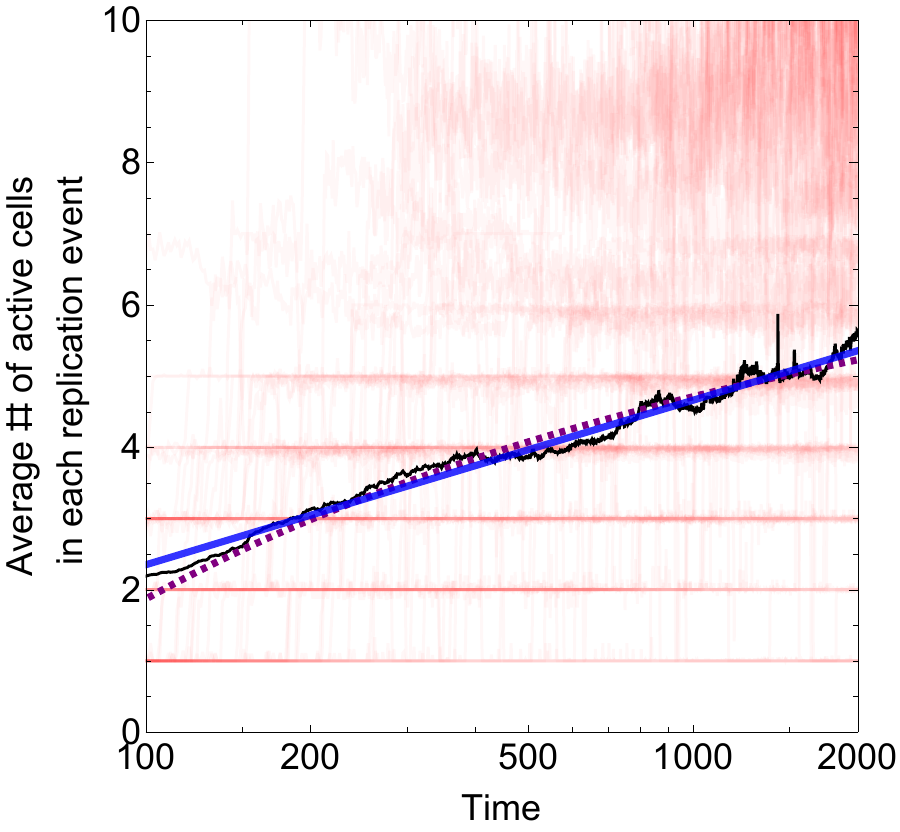}
\caption{Maximum (top) and average (bottom) numbers of active cells in replicating patterns. The red curves show results of 87 independent simulation runs, while the black solid curve shows their average. The time is in log scale to show long-term trends clearly. Purple (dashed) and blue (solid) curves are two different growth models (purple: bounded growth, blue: unbounded growth) fitted to the average behaviors during the time period 100--2000. In both plots, the unbounded growth model (blue curve) was a significantly better fit. See Table \ref{tab:curvefits} for more details.}
\label{fig:higher-order}
\end{figure}

\begin{figure}[t!]
\centering
\includegraphics[width=\columnwidth]{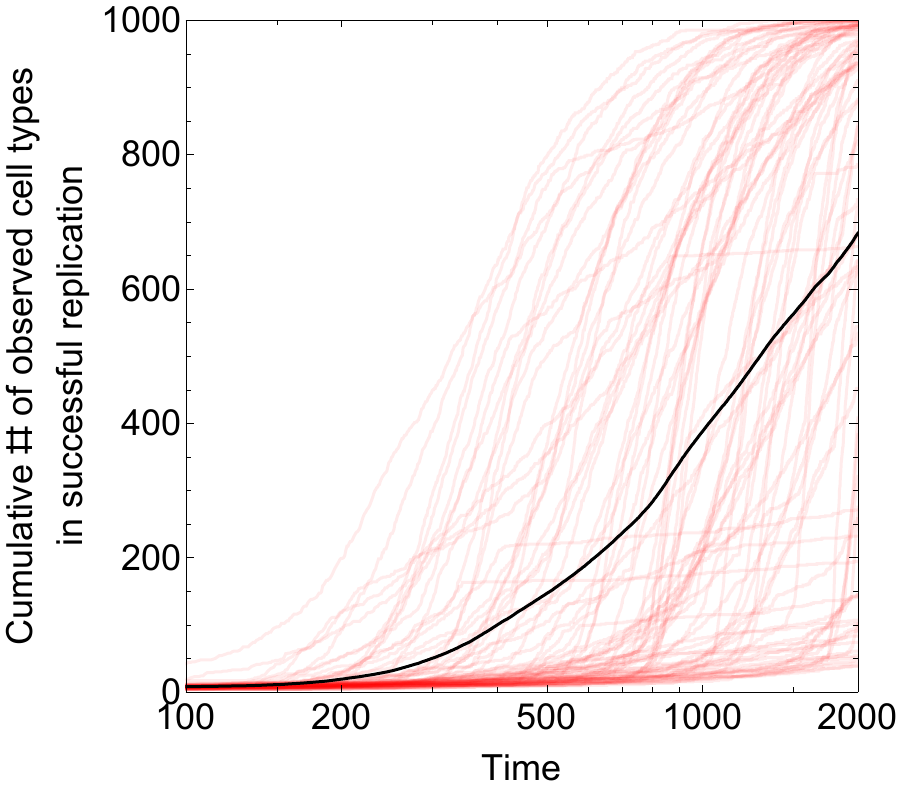}
\includegraphics[width=\columnwidth]{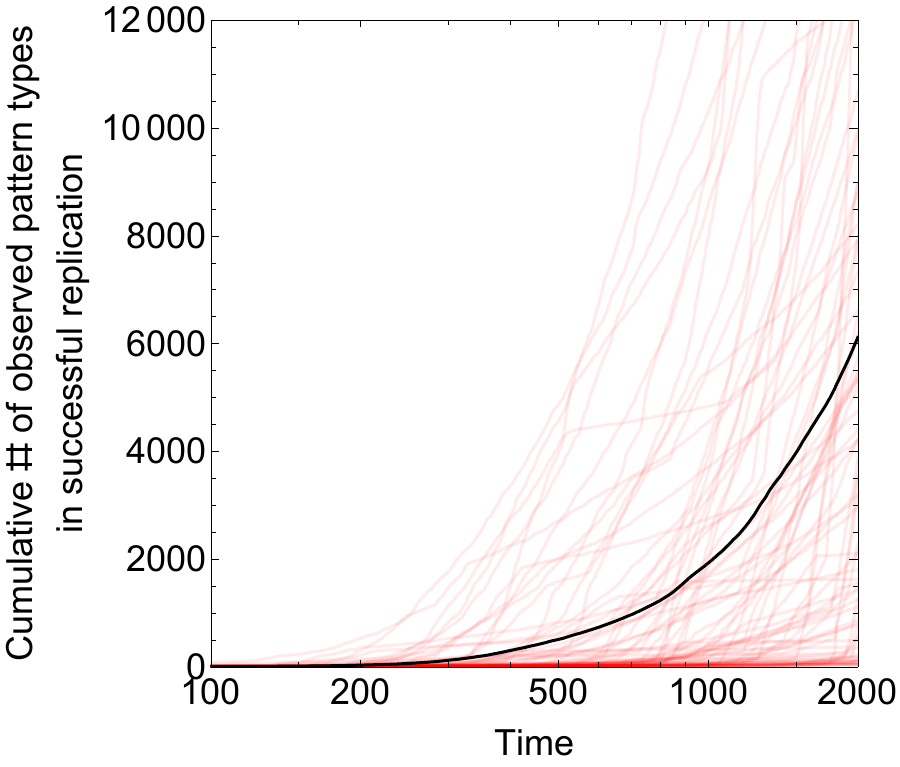}
\caption{Cumulative numbers of observed unique types that appeared in successful replication in the course of simulation. Top: Cumulative number of observed cell types. Bottom: Cumulative number of observed pattern (connected component) types. The red curves show results of 87 independent simulation runs, while the black solid curve shows their average. The time is again in log scale. Note that the curves in the top plot saturate at the maximal number of types ($k=1000$), whereas the curves in the bottom plot grow toward substantially large numbers without bound.}
\label{fig:open-ended}
\end{figure}

Figure \ref{fig:higher-order} shows the evolution of the average size of replicating patterns in SCHC. It was clearly observed in these plots that the dominant patterns became larger over time, demonstrating the spontaneous evolution of higher-order structures that was also observed in the original and non-spatial Hash Chemistry models \cite{sayama2019,sayama2024a}. Following these earlier studies, we fitted two different growth models (bounded and unbounded) to the average pattern size behaviors for $t=100-2000$ in logarithmic time scales. Details of the models and the results are shown in Table \ref{tab:curvefits}. In both the maximum number (Fig.\ \ref{fig:higher-order}, top) and the average number (Fig.\ \ref{fig:higher-order}, bottom) of active cells, the unbounded growth model was a significantly better fit to the data (see Table \ref{tab:curvefits}; compare the AIC/BIC values between bounded and unbounded growth models). This result successfully demonstrates that the unbounded growth of higher-order entity size can be achievable in spatial Hash Chemistry models even in the presence of spatial pressure. 

\begin{table*}[t!]
\centering
\caption{Summary of curve fitting of two different growth models to the results shown in Fig.\ \ref{fig:higher-order}}
\begin{tabular}{llll}
\hline
\multicolumn{1}{c}{Data} & \multicolumn{1}{c}{Model} & \multicolumn{2}{c}{Results} \\
\hline
Maximum number of active cells & Bounded growth &  \multicolumn{2}{l}{Best fit:} \\
in replicating patterns (Fig.\ \ref{fig:higher-order}, top) & $n(t) = -a / \log t + b$ & 
\multicolumn{2}{l}{$n(t) = -105.785 / \log t + 23.2484$} \\
 & & $R^2$ & $0.993901$\\
 & & AIC & $3447.74$ \\
 & & BIC & $3464.39$ \\
 \cline{2-4}
 & Unbounded growth &  \multicolumn{2}{l}{Best fit:} \\
 & $n(t) = a \log t + b$ & 
\multicolumn{2}{l}{$n(t) = 2.73518 \log t - 11.0874$} \\
 & & $R^2$ & $0.996233$\\
 & & AIC & $2532.15$ \\
 & & BIC & $2548.80$ \\
\hline
Average number of active cells & Bounded growth &  \multicolumn{2}{l}{Best fit:} \\
in replicating patterns (Fig.\ \ref{fig:higher-order}, bottom) & $n(t) = -a / \log t + b$ & 
\multicolumn{2}{l}{$n(t) = -39.2514 / \log t + 10.3927$} \\
 & & $R^2$ & $0.998872$\\
 & & AIC & $-1724.65$ \\
 & & BIC & $-1708.00$ \\
 \cline{2-4}
 & Unbounded growth &  \multicolumn{2}{l}{Best fit:} \\
 & $n(t) = a \log t + b$ & 
\multicolumn{2}{l}{$n(t) = 1.00364 \log t - 2.27157$} \\
 & & $R^2$ & $0.99923$\\
 & & AIC & $-2449.38$ \\
 & & BIC & $-2432.73$ \\ 
\hline
\end{tabular}
\label{tab:curvefits}
\end{table*}

\begin{table*}[t!]
\centering
\caption{Comparison of four Hash Chemistry models}
\begin{tabular}{lcccc}
\hline
Attributes & Original \cite{sayama2019} & Non-spatial \cite{sayama2024a} & Cellular (prototype) \cite{sayama2024b} & SCHC (this study) \\
\hline
Selection mechanism & Individual evaluation & Pairwise competition & Four-way competition & Pairwise competition \\
Unbounded possibilities by cardinality leap & \checkmark & \checkmark & \checkmark & \checkmark \\
Computational efficiency & & \checkmark & \checkmark & \checkmark \\
Multiscale ecological interactions & \checkmark & & \checkmark & \checkmark \\
Adaptive behavior & ? & \checkmark & & \checkmark \\
Continuous exploration of possibility space & \checkmark & ? & \checkmark & \checkmark \\
Unbounded complexity growth & ? & \checkmark & & \checkmark \\
\hline
\end{tabular}
\label{model-comparison}
\end{table*}

Finally, Figure \ref{fig:open-ended} shows the cumulative numbers of observed unique types that appeared in successful replication of patterns in the course of simulation run (top: observed cell types; bottom: observed pattern types). It was seen that the cumulative number of unique active cell types quickly increased and could saturate at the maximum value $k=1000$, whereas the cumulative number of unique higher-order pattern types kept increasing to tens of thousands without bound. Such a clear difference in the numbers of possibilities between individual-level and higher-order entities was apparent in the original Hash Chemistry model \cite{sayama2019} but it was much weaker in the non-spatial version because of the overly strong selection pressure and loss of diversity in non-spatial well-mixed settings \cite{sayama2024a}. Figure \ref{fig:open-ended} shows that the proposed SCHC model successfully recovered the diversity in evolutionary populations and continuous evolutionary exploration of the possibility space, providing clear evidence of the effects of the cardinality leap in spatial Hash Chemistry models. 

\section{Conclusions}

In this paper, we proposed Structural Cellular Hash Chemistry (SCHC), a novel variant of Cellular Hash Chemistry in which higher-order entities were represented as connected components in a nearest neighbor graph of active (non-empty) cells in a 2D grid space. The model assumptions and the simulation algorithm of SCHC were carefully designed so that they would preserve the individual identities of connected components while leaving sufficient room for evolutionary exploration and adaptation through mutations and spatial interactions. Numerical simulation experiments demonstrated that SCHC successfully achieved all the desired attributes sought for minimalistic open-ended evolutionary systems (see Table \ref{model-comparison} for comparison of four models), including computational efficiency, multiscale ecological interactions, and unbounded complexity growth of higher-order structures.

The successful implementation of SCHC proves through the construction of a concrete example that it is possible to have a simple Artificial Chemistry system in which both multiscale ecological interactions of self-replicators and their unbounded complexity growth can be simulated at once without prohibitive computational costs. We believe one of the key aspects of SCHC is to use connected components in a nearest neighbor graph on a regular spatial grid as the representation of higher-order entities, for which computationally efficient algorithms exist. We consider SCHC as a major milestone in a series of recent Hash Chemistry research, achieving multiple desired attributes in a single, straightforward model framework.

There are many directions for future research from here. Given the computational efficiency of SCHC, one can easily scale up the space at multiple orders of magnitude using GPUs and/or HPC infrastructure to explore what kind of evolutionary dynamics emerge. Another interesting direction is to conduct more in-depth phylogenetic analysis of evolved patterns (e.g., Fig.\ \ref{size-growth-example}) to evaluate how open-ended this model really is. Finally, one can replace the hash function with any other computational function that can compute a consistent value for a spatial pattern of any size/complexity, to examine what kind of functions will be most useful/insightful/entertaining. 

\newpage
\bibliographystyle{IEEEtran}
\bibliography{IEEE-SSCI-2025}

\begin{thebibliography}{10}
\providecommand{\url}[1]{#1}
\csname url@samestyle\endcsname
\providecommand{\newblock}{\relax}
\providecommand{\bibinfo}[2]{#2}
\providecommand{\BIBentrySTDinterwordspacing}{\spaceskip=0pt\relax}
\providecommand{\BIBentryALTinterwordstretchfactor}{4}
\providecommand{\BIBentryALTinterwordspacing}{\spaceskip=\fontdimen2\font plus
\BIBentryALTinterwordstretchfactor\fontdimen3\font minus
  \fontdimen4\font\relax}
\providecommand{\BIBforeignlanguage}[2]{{%
\expandafter\ifx\csname l@#1\endcsname\relax
\typeout{** WARNING: IEEEtran.bst: No hyphenation pattern has been}%
\typeout{** loaded for the language `#1'. Using the pattern for}%
\typeout{** the default language instead.}%
\else
\language=\csname l@#1\endcsname
\fi
#2}}
\providecommand{\BIBdecl}{\relax}
\BIBdecl

\bibitem{stanley2019open}
K.~O. Stanley, ``Why open-endedness matters,'' \emph{Artificial Life}, vol.~25,
  no.~3, pp. 232--235, 2019.

\bibitem{packard2019overview}
N.~Packard, M.~A. Bedau, A.~Channon, T.~Ikegami, S.~Rasmussen, K.~O. Stanley,
  and T.~Taylor, ``An overview of open-ended evolution: Editorial introduction
  to the open-ended evolution {II} special issue,'' \emph{Artificial Life},
  vol.~25, no.~2, pp. 93--103, 2019.

\bibitem{stepney2021modelling}
S.~Stepney, ``Modelling and measuring open-endedness,'' \emph{Artificial Life},
  vol.~25, no.~1, p.~9, 2021.

\bibitem{borg2023evolved}
J.~M. Borg, A.~Buskell, R.~Kapitany, S.~T. Powers, E.~Reindl, and C.~Tennie,
  ``Evolved open-endedness in cultural evolution: A new dimension in open-ended
  evolution research,'' \emph{Artificial Life}, pp. 1--22, 2023.

\bibitem{stepney2023open}
S.~Stepney and S.~Hickinbotham, ``On the open-endedness of detecting
  open-endedness,'' \emph{Artificial Life}, pp. 1--26, 2023.

\bibitem{adams2017}
A.~Adams, H.~Zenil, P.~C. Davies, and S.~I. Walker, ``Formal definitions of
  unbounded evolution and innovation reveal universal mechanisms for open-ended
  evolution in dynamical systems,'' \emph{Scientific Reports}, vol.~7, no.~1,
  p. 997, 2017.

\bibitem{chan2023}
B.~W.-C. Chan, ``Towards large-scale simulations of open-ended evolution in
  continuous cellular automata,'' in \emph{Proceedings of the Companion
  Conference on Genetic and Evolutionary Computation}, 2023, pp. 127--130.

\bibitem{nichele2024special}
S.~Nichele, H.~Sayama, E.~Medvet, C.~Nehaniv, and M.~Pavone, ``Editorial:
  Special issue “the distributed ghost”—cellular automata, distributed
  dynamical systems, and their applications to intelligence,'' pp. 1--3, 2024.

\bibitem{sayama2024review}
\BIBentryALTinterwordspacing
H.~Sayama and C.~L. Nehaniv, ``Self-reproduction and evolution in cellular
  automata: 25 years after evoloops,'' \emph{Artificial Life}, pp. 1--15, 2024.
  [Online]. Available: \url{https://doi.org/10.1162/artl\_a\_00451}
\BIBentrySTDinterwordspacing

\bibitem{sayama2019}
H.~Sayama, ``Cardinality leap for open-ended evolution: Theoretical
  consideration and demonstration by {Hash} {Chemistry},'' \emph{Artificial
  Life}, vol.~25, no.~2, pp. 104--116, 2019.

\bibitem{sayama2024a}
------, ``Non-spatial {Hash Chemistry} as a minimalistic open-ended
  evolutionary system,'' in \emph{Proceedings of the 2024 IEEE Congress on
  Evolutionary Computation (CEC 2024)}.\hskip 1em plus 0.5em minus 0.4em\relax
  IEEE, 2024, in press; preprint available at
  \url{https://arxiv.org/abs/2404.18027}.

\bibitem{sayama2024b}
------, ``{Hash Chemistry} on a cellular grid: An open-ended artificial
  chemistry system with computational efficiency and nontrivial spatio-temporal
  dynamics,'' in \emph{Artificial Life Conference Proceedings}.\hskip 1em plus
  0.5em minus 0.4em\relax MIT Press, 2024, pp. Paper No: isal\_a\_00\,762, 116.

\end{thebibliography}

\end{document}